# Vision Language Models for Radiation Patterns to Antenna Parameters

Pallaviram Sure[1] and Chandra Mohan Bhuma[2]
[1]Department of Electronics and Communication Engineering, M S Ramaiah University of Applied Sciences, India
[2]Department of Electronics and Communication Engineering, Bapatla Engineering College, India

***Abstract*— Observed radiation patterns often serve as a primary evidence of antenna's behavior, but translating them into meaningful interpretations is a nontrivial and expertise-intensive task. This demand necessitates automated pattern interpretation, a diagnosis problem encountered in applications encompassing Radio Frequency (RF) surveillance, non-cooperative emitter characterization and Over-The-Air (OTA) testing. This work addresses the incorporation of Contrastive Language-Image Pre-training (CLIP) and other recent vision-language models to perform such a diagnosis. These models analyze the multimodal data and extract discriminative features directly from radiation pattern images, to train Machine Learning (ML) approaches for meaningful inferences on geometrical and performance parameters of the antenna. Performance of three such vision language models is demonstrated using RadPat-50K, a synthetic dataset of radiation pattern images generated for uniform linear arrays (ULAs). The radiation pattern images are processed by the vision language models to obtain discriminative features. The CLIP-extracted features are processed by either ML classification models to infer antenna parameters—number of elements, element spacing, weighting scheme, presence of grating lobe and steering angle, or ML regression models to infer parameters—beam width, directivity and main lobe direction. ML models achieve accuracies exceeding 70%, highlighting the potential of multimodal Artificial Intelligence (AI) towards intuitive antenna analysis.**

## I. Introduction

Radiation pattern is an indicative graphical measure of the antenna's performance in terms of directivity, beam width, side lobe levels, and null positions, which are crucial to understand antenna's relative power density behavior. In case of antenna arrays with identical elements, the radiation pattern is proportional to the array factor, which changes as a function of antenna's geometrical parameters - number of elements in the array, element spacing, steering angle, and weighing scheme. Considering a particular radiation pattern, reverse inference of antenna array's geometrical parameters and performance parameters is a diagnosis problem, often helpful in applications like OTA, RF surveillance and non-cooperative emitter characterization.

OTA testing helps in evaluating antenna and also the wireless system performance in realistic operating conditions, particularly for phased-array and mmWave systems (Pei 2019). In OTA chambers, devices under test are assessed primarily through their radiated fields, as there is no direct access to the device's internal antenna geometries and feed networks. Hence radiation patterns are a crucial artifact for diagnosing antenna behavior, while also verifying design intent, and identifying deviations caused by manufacturing tolerances, hardware faults, or calibration errors. In scenarios of RF surveillance and electronic intelligence, emitters are typically non-cooperative, encrypted, or undocumented, direct knowledge of antenna's geometrical parameters are completely unavailable. In such environments, observed radiation patterns are the only exploitable signatures for assessing emitter capabilities and operational intent (Zhang et al. 2022). Information such as beam width, directivity, side lobe structure, and steering provides critical cues for classifying emitters (Liu et al. 2022), distinguishing communication systems from radars or jammers, and inferring array sophistication.

In all the above mentioned applications, extracting geometrical and performance parameters from measured or intercepted radiation patterns are of huge interest. Traditionally, these parameters are obtained through analytical computation or numerical simulation of the array factor. However, such methods can be computationally intensive, especially when exploring large design spaces with multiple variations in element number, spacing, and weighting schemes. This motivates the development of data-driven approaches to infer antenna parameters directly from radiation pattern images which is the primary motto of this work.

Specifically, this work addresses an automated radiation pattern interpretation by formulating the problem as a diagnostic analysis or a reverse inference problem. With the aid of CLIP and vision language models, discriminative features are extracted from radiation pattern images, which are subsequently used by ML classification and regression models to infer antenna's geometrical or performance parameters. The performance of these models is tested on RadPat-50K, a large-scale synthetic dataset with simulated radiation patterns of ULAs (Chandra et al. 2025) having a total of 50000 radiation pattern images.

The ULA considered in the dataset consists of antenna elements arranged along a straight line with equal spacing between adjacent elements. Each element is typically identical, and the array factor is determined by the number of elements, their relative spacing in terms of wavelength, and the excitation weights applied to each element. ULAs are attractive due to their mathematical tractability, ease of implementation, and versatility in beamforming applications (Smith 2023). The radiation pattern of ULA describes the spatial distribution of radiated energy as a function of angle through the array factor.

Key parameters derived from radiation patterns are as follows: Gain is the ability of an antenna to focus energy in a particular direction, which is proportional to directivity mathematically the ratio of main lobe radiation intensity to the average intensity. Beam width corresponds to the angular width of the main lobe, indicating resolution and coverage. Grating lobes are the secondary lobes arising from excessive element spacing, while steering angle is the direction in which the main lobe is oriented (Balanis 2005).

In this work, the ability of CLIP and state-of-the-art vision–language models (VLMs) to extract meaningful features directly from radiation pattern images is evaluated. Results demonstrate that these models can infer critical parameters—including number of elements, element spacing, weight distribution, peak directivity, half power beam width (HPBW), grating lobe presence, and steering angle—without explicit numerical computation. Further we benchmark multiple classification and regression models trained on CLIP-extracted features, consistently achieving accuracies above 70% across different tasks. These results affirm the efficacy of multimodal AI in capturing the latent structure of radiation patterns and highlight the potential of RadPat-50K as a benchmark for advancing AI-driven antenna design. By bridging electromagnetic theory with multimodal learning, this work contributes a scalable dataset and methodological framework that can accelerate innovation in antenna engineering and foster new paradigms in intelligent design automation. Importantly, RadPat-50K is publicly available on Hugging Face and other repositories, supporting reproducibility and community-driven research.

The rest of the manuscript is organized as follows. In section II, the literature survey is presented. The dataset is briefed in section III. The vision language models are summarized in section IV and the parameter inference framework is proposed in section V. The obtained results are discussed in section VI, while the paper is concluded in section VII.

## II. LITERATURE SURVEY

Primary motive of this work is to demonstrate the performance of CLIP and vision language models for diagnosing radiation patterns to obtain antenna parameters that typically aid in RF surveillance and OTA testing. The dataset chosen for demonstration is the RadPat-50k generated with ULAs. The ULAs are a popular category of antennas that witnessed significant progress in the study and application across wireless communication, radar, and satellite systems. Some of these studies are discussed below.

A hybrid design combining uniform and non-uniform linear arrays with circularly polarized patch antennas is optimized for the 5G NR257 band (Lakshman et al. 2025), which rendered higher gain and reduced cross polarization features, making ULAs suitable for mmWave communication. An investigation on near field communications depicted that structural design of ULAs impact spatial multiplexing and energy efficiency in dense Internet of Things (IoT) deployments (Chen et al. 2025). Kepler optimization algorithm is devised and employed for radiation pattern synthesis in ULAs, for side lobe suppression and improved beam shaping, highlighting optimization in the design of antenna arrays (Tang et al. 2025). A comparative study of circular, planar, and linear arrays is conducted for satellite communications (Muttiah 2024) demonstrating ULAs radiation characteristics for 5G applications. ULAs were explored for holographic communication demonstrating that polarization diversity enhances spatial multiplexing capacity (Mestre et al. 2024). MIMO radar systems with ULAs have shown improved bit error rate (BER) performance through adaptive weighting schemes (Lee et al. 2024).

Biogeography Optimization (BO) approaches have also been employed for ULA synthesis. BO Weighed Quantum Wolf Optimization (WQWO) algorithm is studied for improved side lobe control in ULAs (Sahithi and Siddaiah, 2023). The algorithm adjusts excitation magnitudes to suppress side lobes and there by renders better efficiency than traditional BO approaches. A phase locked loop (PLL) based beam steering mechanism is devised for effective phase control and dynamic steering in ULAs (Chepala et al. 2023). A multi verse optimization algorithm is explored for ULA pattern synthesis, to improve beam width and side lobe suppression (Raghuvanshi et al. 2023).

The ULAs have theoretical assumptions and have limitations in array factor modeling (Friedlander 2021), leaving a gap between idealized ULAs and real-world arrays. The work portrays need for more accurate modeling that includes mutual coupling and element patterns. In this regard, ML classification and regression models can be used in the context of several antenna design tasks. Many such models exist which are usually trained on a set of given features and target variables, so as to predict the target for a new set of features.

A review focusing specifically on regression methods in machine learning, providing detailed analysis and comparison of popular regression techniques such as linear regression, polynomial regression, decision tree regression, LASSO, random forest regression, and neural network-based regression (Shahane, 2021) explains the theoretical foundations of these models, their computational properties, advantages and limitations, and discusses how performance varies with dataset characteristics and complexity. The survey serves as a practical guide for selecting regression approaches based on problem structure and predictive requirements.

Another survey work on ML classifiers provides a structured overview of major supervised classification algorithms, including decision trees, kNN, SVM, naïve Bayes, and artificial neural networks (Kotsiantis 2007). The work compares their theoretical foundations and computational characteristics, and discusses evaluation metrics such as accuracy, precision, recall, and cross-validation. Some of the recent works have applied ML approaches to automate antenna design and performance prediction in OTA contexts, such as ML-based prediction of return loss (Jain R 2025) and other performance metrics for IoT and 5G antennas and neural-network-driven optimization of MIMO antenna parameters (Ghasemi M 2025).

Similarly, in RF surveillance and non-cooperative emitter recognition, deep learning models such as domain adversarial neural networks for specific emitter identification (Li D et al. 2024) and ML-based fingerprint extraction frameworks (Zhao Y et al. 2023), motivate the direction of our work on extracting high-level pattern features directly from measured or simulated radiation patterns.

Some of the available literature on the deep learning based approaches for automated interpretation of antenna radiation pattern images is discussed here. A deep neural network that ingests radiation pattern images and predicts the amplitude and phase excitations for each array element, effectively solving the inverse mapping from desired pattern to excitations is proposed (Kim and Choi 2020). Curating 6,859 samples on a 4×1 patch array by sweeping element phases, the work demonstrates better reconstruction of target patterns in the presence of inter-element coupling. By learning nonlinear mappings, the approach bypasses repeated EM solves and generalizes across array configurations, and depicts deep learning models for inverse EM design and pattern-to-parameter interpretation.

Long Short-Term Memory – Recurrent Neural Network (LSTM-RNN) are employed to learn from optimized radiation patterns, and predict complex excitations for ULAs (Arce et al. 2025). Compared with fully connected and convolutional neural network (CNN) baselines, the LSTM exhibits significantly better performance, highlighting the benefits of sequence-aware models in beam synthesis.

A deep learning model that simultaneously controls beam scan angle and side lobe level through radiation pattern profiles predicts element amplitudes and phases (Abdullah et al. 2025). Automating the interpretation of desired patterns into implementable weights, the work demonstrated the fabrication of microstrip arrays. A Bayesian-optimized CNN efficiently predicted far-field radiation patterns of arrays, reducing the need for expensive EM simulations (Zhang et al. 2025). Learning spatial features from pattern data, the CNN model supports rapid design and diagnostics.

The recent advanced vision–language models particularly utilize the CLIP-based architectures. These models have significantly expanded the scope of multimodal AI across diverse application domains. Other than the conventional image classification applications, these models are increasingly employed for complex real-world tasks, some of which include autonomous driving, scene text recognition, and video action understanding.

CLIP is employed for dynamic scene processing and understanding, in autonomous driving environments (Elhenawy et al. 2025). An approach is proposed that intrigues CLIP's zero-shot and prompt-based learning capabilities to classify and interpret complex road scenes. By aligning visual inputs with semantically meaningful textual prompts, the framework improves adaptability to varying traffic conditions and enhances robustness in real-time driving scenarios compared to conventional vision-only systems.

A CLIP-LLaMA which is a hybrid framework combines CLIP's visual encoder with a large language model to address scene text recognition (Zhao et al. 2024). Improved recognition ability is demonstrated in scenarios of irregular, distorted, and low-resolution text that is common in real-world environments. By incorporating strong language priors alongside visual-semantic alignment, the model demonstrated better performance across multiple benchmark datasets.

CLIP is extended to the video domain by introducing video-language prompting and lightweight adaptation strategies for action recognition tasks (Zhang et al. 2025). The method incorporates temporal modeling while preserving CLIP's inherent zero-shot and few-shot capabilities. The approach depicted encouraging results on well-known action recognition benchmark datasets. Hence it depicts that vision–language pre training can be effectively transferred to spatio-temporal learning problems.

Vision–language models such as CLIP are studied for visual recognition tasks (Zhou et al. 2023). The study examines classifier design, feature adaptation, and prompt learning strategies from both theoretical and empirical perspectives. The findings demonstrated improved performance in classification, detection, and segmentation tasks through better usage of multimodal pre-trained representations.

A framework for automatic low probability of intercept (LPI) radar waveform recognition based on a vision–language model is proposed (Yang et al. 2025). The method converts radar time-frequency representations into image-like inputs and aligns them with context prompt embedding in a shared multimodal space. This set-up enables the model to learn intrinsic waveform characteristics and discriminate overlapping signals even under noise, thus aiding radar modulation recognition.

Sig2text, a vision–language model is designed for non-cooperative radar signal parsing and modulation recognition (Feng et al. 2025). The approach uses a vision transformer to extract features from time-frequency radar representations and combines them with transformer-based decoders to identify modulation types and corresponding parameters. Radar signal analysis is casted as a multimodal parsing problem, to effectively interpret complex RF signals, demonstrated using synthetic radar datasets.

Based on all the above, it can be observed that vision language models have been used in many domains but their application to automated interpretation of radiation pattern images radiation pattern images is very much limited, which encourages the work described in this manuscript.

## III. DATASET ATTRIBUTES

RadPat-50K is a large synthetic dataset created to support machine learning research in antenna radiation pattern analysis. It contains 50,000 samples of radiation patterns generated from Uniform Linear Arrays (ULAs). To evaluate model performances, instead of using real antenna measurements—which can be expensive, slow, and limited in diversity—RadPat50K is produced entirely through simulation, allowing full control over the antenna configuration and signal conditions.

Each sample in the dataset represents the radiation behavior of an antenna array under a unique set of parameters. These parameters include the number of antenna elements, the spacing between them, their current amplitudes and phases i.e.

weighting scheme, and the direction toward which the array is steered. By randomly varying these properties across wide and realistic ranges, the dataset captures a rich variety of beam shapes, main lobes, side lobes, nulls, and steering directions. Note that noise can also be added to make the patterns of the dataset to obtain more practical patterns.

Every entry consists of the angle values, the corresponding normalized radiation pattern, and metadata describing how the pattern was generated. RadPat50K is suitable for training and evaluating deep learning models that perform many other tasks such as beam prediction, direction-of-arrival estimation, antenna classification, and inverse modeling. Its size, diversity, and full parameter transparency makes it an excellent benchmark for antenna-related machine learning research.

Each sample of RadPat-50K is generated analytically using array-factor equations under varying geometrical and signal parameters of the antenna array. Consider a ULA with $N$ identical isotropic radiators spaced by $d,$ along the X-axis. The corresponding element positions are given as:

$$x_n = (n-1)d, \quad n = 1,2,\dots,N. \tag{1}$$

For an arrival angle $\theta$, the phase difference between any two adjacent elements, corresponds to

$$\psi(\theta) = \frac{2\pi d cos\theta}{\lambda} \tag{2}$$

where the wavelength $\lambda = \frac{c}{f}$, for a frequency of operation $f$ and speed of light in free space $c$. Array factor becomes

$$AF(\theta) = \sum_{n=1}^{N} w_n exp\big(j(n-1)\psi(\theta)\big) \tag{3}$$

with $w_n$ as the complex weight of the $n^{th}$ element Corresponding power pattern is related as $P(\theta) = |AF(\theta)|^2$, where $\theta \in [-90°, 90°]$. The dataset spans a wide range of configurations, whose geometrical parameters are listed in Table I. Some of the dataset entries comprising the radiation pattern image along with corresponding text geometrical parameters and performance parameters (Grating lobe presence, peak directivity, HPBW and main lobe direction) are shown in Figure 1 for reference.

## IV. Model Description

CLIP, introduced by OpenAI, is a vision–language model that intends to learn joint representations of images and text. It takes an image and a text prompt as input, and encodes them separately using a visual encoder such as vision transformer and a text transformer. Subsequently, CLIP maps both into a shared embedding space and outputs a similarity score between image and text embeddings. CLIP is trained using a contrastive learning objective, where matching image–text pairs are pulled together in embedding space and mismatched pairs are pushed apart. This enables zero-shot classification, image retrieval, and multimodal reasoning without task-specific fine-tuning. Different vision language models used in this work are described below.

### *A. ViT H 14 378 quickgeludfn5b:*

This model is a Vision Transformer (ViT) variant with a large "H" sized backbone (ViT H/14), trained at high resolution (378 pixels) (Apple 2023). It employs the QuickGELU activation function, which accelerates convergence while preserving accuracy. The architecture divides radiation pattern images into 14×14 patches, processes them through transformer layers, and learns global dependencies across the entire image. The model is optimized for robust visual feature extraction, making it suitable for downstream tasks such as classification, retrieval, and multimodal alignment. Its design emphasizes scalability and efficiency, balancing computational cost with high quality representations for large scale vision–language applications.

TABLE I. Geometrical Parameters of The Dataset

| Geometrical Parameter | Values |
|---|---|
| Array elements (N): | 4, 8, 12, 16, 24, 32, 48, 64 |
| Element spacing (λ): | 0.25 λ, 0.5 λ, 0.75 λ, 1.0 λ |
| Steering angles (°): | –60 to +60 in 15° increments |
| Weighting schemes: | Uniform, Binomial, Cosine, Kaiser, Hamming, Hann, Blackman, Exponential |
| Variations: | Amplitude noise, phase noise, steering jitter |

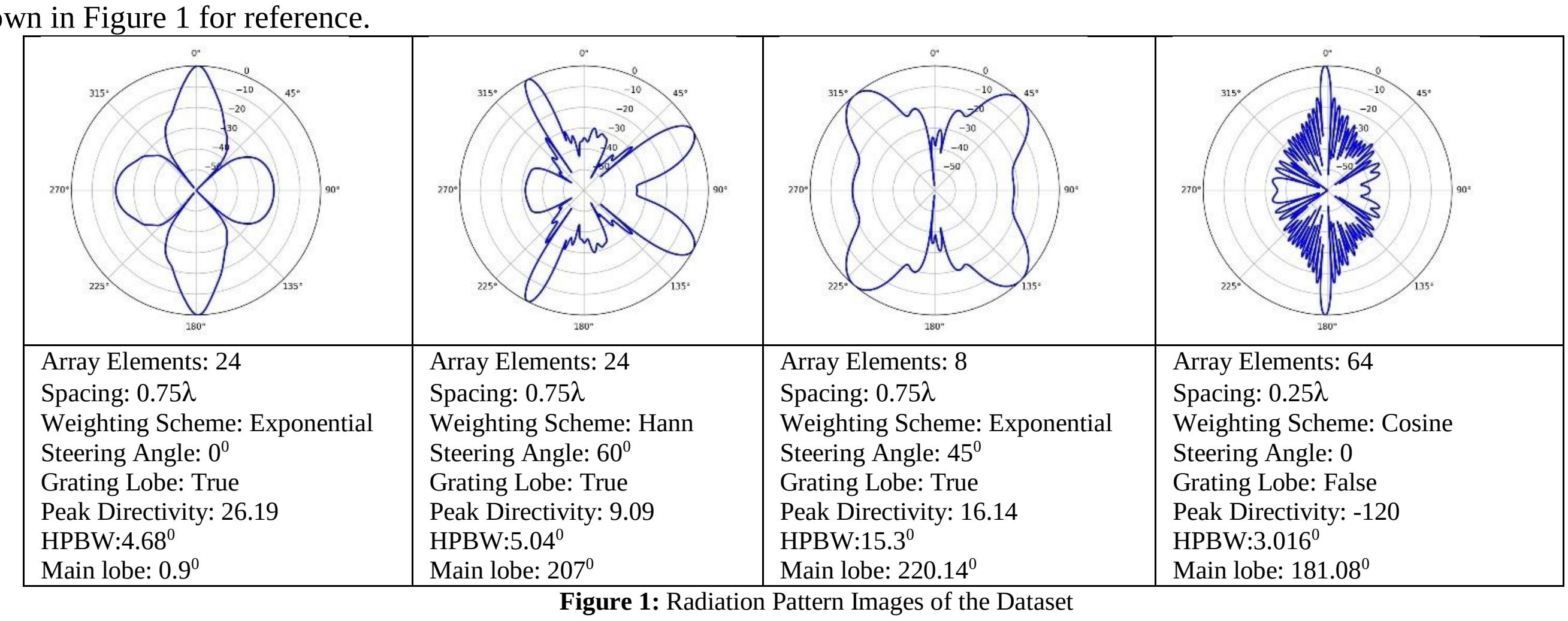

**Figure 1:** Radiation Pattern Images of the Dataset

### B. ViT gopt 16 SigLIP2 384webli:

This model combines a ViT g/16 backbone (a giant Vision Transformer with 16×16 patches) with SigLIP2, a second generation variant of the sigmoid loss for Image–Text pretraining (Zhai X 2023, Google 2025). It is trained at 384 pixel resolution on WebLI, a large multilingual web scale dataset. SigLIP2 improves alignment between visual and textual embeddings by replacing contrastive softmax objectives with sigmoid based losses, yielding stronger cross modal retrieval and captioning performance. The model is designed for multilingual and multimodal tasks, excelling in zero shot transfer across diverse domains. Its large capacity and improved loss function make it a powerful foundation for vision–language reasoning.

### C. DINOv3 ViT 7B 16 pretrain lvd1689m:

This model is part of the DINOv3 family, which extends self supervised learning for Vision Transformers(Caron M 2021, Meta AI 2025). The ViT 7B/16 backbone is extremely large (7 billion parameters) with 16×16 patch embeddings. It is pretrained on the LVD 1689M dataset, a massive curated collection of images. DINOv3 uses self distillation without labels, enabling the model to learn rich semantic features by predicting representations across augmented views. This approach yields highly transferable embeddings for tasks such as detection, segmentation, and retrieval. The scale of ViT 7B combined with DINOv3's self-supervised training renders tranferable visually rich representations.

The above discussed models ViT-H-14-378 (CLIP-based), ViT-gopt-16-SigLIP2-384, and DINOv3-ViT-7B/16 represent advanced transformer-based vision systems designed for large-scale representation learning. The first two are vision–language models, taking both image and text inputs and producing aligned multimodal embeddings for tasks like zero-shot classification and retrieval using contrastive or sigmoid-based objectives. In contrast, DINOv3 is a self-supervised vision-only model that takes images as input and outputs rich visual feature embeddings without language supervision. While CLIP-style and SigLIP models focus on cross-modal alignment between vision and language, DINOv3 emphasizes strong visual representation learning through self-distillation. Together, they illustrate three complementary paradigms: contrastive multimodal learning, sigmoid-based large-scale alignment, and self-supervised visual pre training.

## V. Proposed Parameter Inference Framework

In this work we propose a multimodal learning framework for radiation pattern analysis using the RadPat-50K dataset. The framework integrates ViT–based encoders with downstream classifiers and regressors to infer antenna parameters directly from radiation pattern images, described in the following steps:

**Step 1-** Dataset Representation: The dataset be denoted as:

$$\mathcal{D} = \{(x_i, y_i, z_i)\}_{i=1}^{N} \tag{4}$$

where $x_i \in \mathbb{R}^{H\times W\times C}$ is radiation pattern image (polar plot), $y_i = (n_i, d_i, w_i)$ which are categorical labels representing (number of elements, spacing, weighting scheme) respectively and $z_i \in \mathbb{R}^4$: continuous value parameters which are (HPBW, steering angle $\theta_s$, peak directivity $D_{peak}$, main lobe direction $\theta_m$).

**Step 2 -** Feature Extraction via Vision Transformers: The three pre-trained encoders used for feature extraction are
M1: ViT-H-14-378-quickgeludfn5b (14 by 14 patches) (IV.*A*)
M2: ViT-gopt-16-SigLIP2-384webli (16 by 16 patches) (IV.*B*)
M3: DINOv3-ViT-7B-16-pretrain-lvd1689m, (IV.*C*)
These models were selected for their strong representation capabilities and demonstrated robustness in capturing fine-grained structures within antenna radiation patterns. Each encoder maps the image $x_i$ into a high-dimensional embedding:

$$f_k: \mathbb{R}^{H\times W\times C} \rightarrow \mathbb{R}^d \tag{5}$$

where $k$ indexes the three feature extractors and corresponding embedding vector from extractor $k$ is denoted as:

$$e_i = f_k(x_i) \in \mathbb{R}^{d_k} \tag{6}$$

with $d_1 = 1024, d_2 = 1536$ and $d_3 = 4096$ corresponding to M1, M2, and M3 respectively.

**Step 3-** Train/Test Split: The dataset $\mathcal{D}$ is partitioned into training and testing subsets $\mathcal{D}_{train}$ and $\mathcal{D}_{test}$ such that

$$\begin{gathered} \mathcal{D}_{train} \cup \mathcal{D}_{test} = \mathcal{D} \\ \mathcal{D}_{train} \cap \mathcal{D}_{test} = \emptyset \\ |\, \mathcal{D}_{train} \,| = 0.8N \\ |\, \mathcal{D}_{test} \,| = 0.2N \end{gathered} \tag{7}$$

where $|(.)|$ represents the cardinality of the set. The embeddings are extracted separately as

$$\begin{gathered} E_{train} = \{e_i \mid (x_i, y_i, z_i) \in \mathcal{D}_{train}\} \\ E_{test} = \{e_j \mid (x_j, y_j, z_j) \in \mathcal{D}_{test}\} \end{gathered} \tag{8}$$

**Step 4a -** Classification Task: A ML classifier $g_c$ is trained on $E_{train}$ by minimizing a classification loss between model outputs and true labels. The trained classifier is then evaluated on the held-out test set to predict categorical antenna parameters as

$$\hat{y}_j = g_c(e_j), \quad e_j \in E_{test} \tag{9}$$

**Step 4b -** Regression Task: A regressor $g_r$ is trained on $E_{train}$ by minimizing a regression loss between predicted and true values. The trained regressor is then evaluated on the held-out test to predict continuous valued antenna parameters as:

$$\hat{z}_j = g_r(e_j), \quad e_j \in E_{test} \tag{10}$$

Note that based on the target variable, the task in Step 4 is performed. If the target variable is $y_i$ classification task in Step

4a is performed and if the target variable is $z_i$, then regression task in Step 4b is performed.

## VI. Results and Discussion

The simulations were conducted using Google Colab Pro, leveraging NVIDIA A100 and L4 GPUs to efficiently process the RadPat-50K dataset. All radiation pattern images were first converted into polar plot images and subsequently encoded into high-dimensional feature embeddings using three state-of-the-art vision transformer models: M1, M2 and M3.

The extracted embeddings were used as input features for downstream learning tasks implemented using the cuML library, enabling GPU-accelerated evaluation of multiple classifiers and regressors. Models such as cuML's Random Forest, Support Vector Machine (SVM) variants namely SVM with Radial Basis Function (RBF) kernel and SVM with Linear kernel, Logistic Regression, k Nearest Neighbor (kNN), and Gaussian Naïve Bayes (NB) were assessed for classification and parameter-prediction performance.

The incorporation of GPU based pipelines significantly reduced feature extraction and model-training time, allowing systematic evaluation across all RadPat-50K data samples. Comparative analysis revealed that transformer-based embeddings consistently produced stable decision boundaries across classifiers.

### A. *Results of Classification*

The classification results with feature extraction via M1, M2 and M3 models on the dataset Polar50K.mat are performed with feature vector sizes of (50000, $d_1$), (50000, $d_2$) and (50000, $d_3$) respectively, where 50,000 corresponds to the number of samples, and $d_1, d_2, d_3$ are the respective embedding dimensions output by M1, M2, and M3 models. The results are shown in Table II to Table VI for grating lobe, number of elements, spacing, weighting scheme and steering angle parameters respectively. Note that these are all multi-level classifications and are achieved using Step 4a of section IV. D.

The performance of the classifiers are given in these Tables in terms of classification accuracy, precision, recall and the F1-score. The train and test set cardinalities are in accordance with (7). Among the tested models, SVM (RBF) showed better performance in predicting majority of the parameters. From the results of **grating lobe presence parameter** in Table II, SVM (RBF) rendered 99.12% accuracy with M1 and 99.44% with M3 extractors. M2 lags behind the other two extractors across every classifier. M2 with KNN has 90.4% accuracy while M2 with SVM (RBF) provided only 87%. Accuracy of Gaussian NB is only 59.9% with M2.

From the results of **number of elements parameter** shown in Table III, M1 with SVM (RBF) gave an accuracy of 99.12% — far ahead of M2 with SVM Linear having 58% and M3 with SVM (RBF) having 76% accuracy. Like in grating lobe case, Gaussian NB with M2 could not cope up with other models, and has as least as 19.5% accuracy.

The results for **spacing parameter** in Table IV show similar inferences as those of grating lobe case. M1 with SVM (RBF) provided 94.55%, while M3 with SVM (RBF) provided 95.93%. M2 favors SVM (Linear) with 71.78% at its best while Gaussian NB with M2 has its least as 32.6%.

TABLE II. Classification Results of Grating Lobe

| | ViT-H-14-378-quickgeludfn5b | | | | ViT-gopt-16-SigLIP2-384webli | | | | dinov3-vit7b16-pretrain-lvd1689m | | | |
|---|---|---|---|---|---|---|---|---|---|---|---|---|
| **Grating Lobe** | **Accuracy** | **Precision** | **Recall** | **F1 Score** | **Accuracy** | **Precision** | **Recall** | **F1 Score** | **Accuracy** | **Precision** | **Recall** | **F1 Score** |
| Random Forest | 0.9699 | 0.969980 | 0.969 | 0.9698 | 0.8896 | 0.8905 | 0.8896 | 0.8895 | 0.9820 | 0.9820 | 0.9820 | 0.9820 |
| SVM (RBF) | **0.9912** | **0.991206** | **0.991** | **0.9912** | 0.8699 | 0.8709 | 0.8699 | 0.8698 | **0.9944** | **0.9944** | **0.9944** | **0.9944** |
| SVM (Linear) | 0.9870 | 0.987001 | 0.987 | 0.9870 | 0.8941 | 0.8942 | 0.8941 | 0.8941 | 0.9929 | 0.9929 | 0.9929 | 0.9929 |
| Logistic Regression | 0.9867 | 0.986700 | 0.986 | 0.9867 | 0.8865 | 0.8867 | 0.8865 | 0.8865 | 0.9939 | 0.9939 | 0.9939 | 0.9939 |
| KNN | 0.9802 | 0.980256 | 0.980 | 0.9802 | **0.9038** | **0.9038** | **0.9038** | **0.9038** | 0.9864 | 0.9864 | 0.9864 | 0.9864 |
| Gaussian NB | 0.7426 | 0.747825 | 0.743 | 0.7412 | 0.5989 | 0.6470 | 0.5994 | 0.5637 | 0.7539 | 0.7552 | 0.7538 | 0.7536 |

TABLE III. Classification Results of Number Of Elements

| | ViT-H-14-378-quickgeludfn5b | | | | ViT-gopt-16-SigLIP2-384webli | | | | dinov3-vit7b16-pretrain-lvd1689m | | | |
|---|---|---|---|---|---|---|---|---|---|---|---|---|
| **No. of Elements** | **Accuracy** | **Precision** | **Recall** | **F1 Score** | **Accuracy** | **Precision** | **Recall** | **F1 Score** | **Accuracy** | **Precision** | **Recall** | **F1 Score** |
| Random Forest | 0.9699 | 0.969980 | 0.9699 | 0.9698 | 0.4978 | 0.5091 | 0.4977 | 0.4972 | 0.6647 | 0.6658 | 0.6645 | 0.6624 |
| SVM (RBF) | **0.9912** | **0.991206** | **0.9912** | **0.9912** | 0.4825 | 0.5277 | 0.4824 | 0.4882 | **0.7608** | **0.7618** | **0.7607** | **0.7594** |
| SVM (Linear) | 0.9870 | 0.987001 | 0.9870 | 0.9870 | **0.5815** | **0.5974** | **0.5814** | **0.5831** | 0.7253 | 0.7272 | 0.7252 | 0.7258 |
| Logistic Regression | 0.9867 | 0.986700 | 0.9867 | 0.9867 | 0.5348 | 0.5440 | 0.5347 | 0.5350 | 0.7217 | 0.7218 | 0.7216 | 0.7216 |
| KNN | 0.9802 | 0.980256 | 0.9802 | 0.9802 | 0.4856 | 0.4765 | 0.4856 | 0.4776 | 0.6702 | 0.6666 | 0.6701 | 0.6666 |
| Gaussian NB | 0.7426 | 0.747825 | 0.7427 | 0.7412 | 0.1950 | 0.1537 | 0.1952 | 0.1404 | 0.3658 | 0.3350 | 0.3656 | 0.3442 |

TABLE IV. CLASSIFICATION RESULTS OF SPACING

| Spacing | ViT-H-14-378-quickgeludfn5b Accuracy | Precision | Recall | F1 Score | ViT-gopt-16-SigLIP2-384webli Accuracy | Precision | Recall | F1 Score | dinov3-vit7b16-pretrain-lvd1689m Accuracy | Precision | Recall | F1 Score |
|---|---|---|---|---|---|---|---|---|---|---|---|---|
| Random Forest | 0.8766 | 0.8777 | 0.8765 | 0.8766 | 0.6851 | 0.6926 | 0.6849 | 0.6824 | 0.9150 | 0.9170 | 0.9149 | 0.9152 |
| SVM (RBF) | **0.9455** | **0.9454** | **0.9454** | **0.9454** | 0.6107 | 0.6130 | 0.6105 | 0.6090 | **0.9593** | **0.9593** | **0.9592** | **0.9593** |
| SVM (Linear) | 0.9304 | 0.9302 | 0.9303 | 0.9302 | **0.7178** | **0.7160** | **0.7176** | **0.7166** | 0.9514 | 0.9514 | 0.9513 | 0.9514 |
| Logistic Regression | 0.9276 | 0.9273 | 0.9275 | 0.9274 | 0.6579 | 0.6551 | 0.6577 | 0.6558 | 0.9534 | 0.9535 | 0.9533 | 0.9534 |
| KNN | 0.8898 | 0.8905 | 0.8896 | 0.8898 | 0.6870 | 0.6866 | 0.6869 | 0.6861 | 0.9201 | 0.9207 | 0.9200 | 0.9203 |
| Gaussian NB | 0.4565 | 0.4552 | 0.4563 | 0.4453 | 0.3260 | 0.3461 | 0.3255 | 0.2468 | 0.4987 | 0.4927 | 0.4982 | 0.4801 |

TABLE V. CLASSIFICATION RESULTS OF WEIGHTING SCHEME

| Weighting Scheme | ViT-H-14-378-quickgeludfn5b Accuracy | Precision | Recall | F1 Score | ViT-gopt-16-SigLIP2-384webli Accuracy | Precision | Recall | F1 Score | dinov3-vit7b16-pretrain-lvd1689m Accuracy | Precision | Recall | F1 Score |
|---|---|---|---|---|---|---|---|---|---|---|---|---|
| Random Forest | 0.5793 | 0.5690 | 0.5796 | 0.5605 | 0.4592 | 0.4555 | 0.4596 | 0.4440 | 0.6209 | 0.6169 | 0.6214 | 0.6049 |
| SVM (RBF) | **0.6906** | **0.6841** | **0.6908** | **0.6789** | 0.4157 | 0.4250 | 0.4161 | 0.3989 | 0.7401 | 0.7375 | 0.7403 | 0.7332 |
| SVM (Linear) | 0.6912 | 0.6846 | 0.6913 | 0.6854 | **0.4953** | **0.4842** | **0.4960** | **0.4803** | 0.7419 | **0.7478** | 0.7417 | 0.7443 |
| Logistic Regression | 0.6828 | 0.6733 | 0.6828 | 0.6754 | 0.4662 | 0.4541 | 0.4669 | 0.4514 | **0.7479** | 0.7469 | **0.7477** | **0.7472** |
| KNN | 0.5311 | 0.5265 | 0.5313 | 0.5244 | 0.4580 | 0.4501 | 0.4581 | 0.4505 | 0.5753 | 0.5699 | 0.5754 | 0.5683 |
| Gaussian NB | 0.3861 | 0.3728 | 0.3870 | 0.3541 | 0.2548 | 0.2456 | 0.2554 | 0.2169 | 0.4232 | 0.4047 | 0.4238 | 0.3969 |

TABLE VI. CLASSIFICATION RESULTS OF STEERING ANGLE (NOMINAL)

| Steering Angle (Nominal) | ViT-H-14-378-quickgeludfn5b Accuracy | Precision | Recall | F1 Score | ViT-gopt-16-SigLIP2-384webli Accuracy | Precision | Recall | F1 Score | dinov3-vit7b16-pretrain-lvd1689m Accuracy | Precision | Recall | F1 Score |
|---|---|---|---|---|---|---|---|---|---|---|---|---|
| Random Forest | 0.7827 | 0.7816 | 0.7827 | 0.7811 | 0.4935 | 0.4934 | 0.4934 | 0.4890 | 0.8488 | 0.8492 | 0.8487 | 0.8487 |
| SVM (RBF) | **0.9119** | **0.9115** | **0.9118** | **0.9114** | 0.3839 | 0.3912 | 0.3839 | 0.3754 | **0.9214** | **0.9207** | **0.9212** | **0.9207** |
| SVM (Linear) | 0.8925 | 0.8928 | 0.8924 | 0.8925 | **0.7670** | **0.7668** | **0.7669** | **0.7667** | 0.9161 | 0.9161 | 0.9159 | 0.9160 |
| Logistic Regression | 0.8794 | 0.8794 | 0.8793 | 0.8793 | 0.6582 | 0.6570 | 0.6582 | 0.6571 | 0.9184 | 0.9181 | 0.9182 | 0.9181 |
| KNN | 0.7829 | 0.7861 | 0.7827 | 0.7829 | 0.4691 | 0.4823 | 0.4688 | 0.4709 | 0.8270 | 0.8281 | 0.8268 | 0.8270 |
| Gaussian NB | 0.3599 | 0.3711 | 0.3607 | 0.3374 | 0.1115 | 0.0661 | 0.1119 | 0.0507 | 0.5201 | 0.5245 | 0.5203 | 0.5127 |

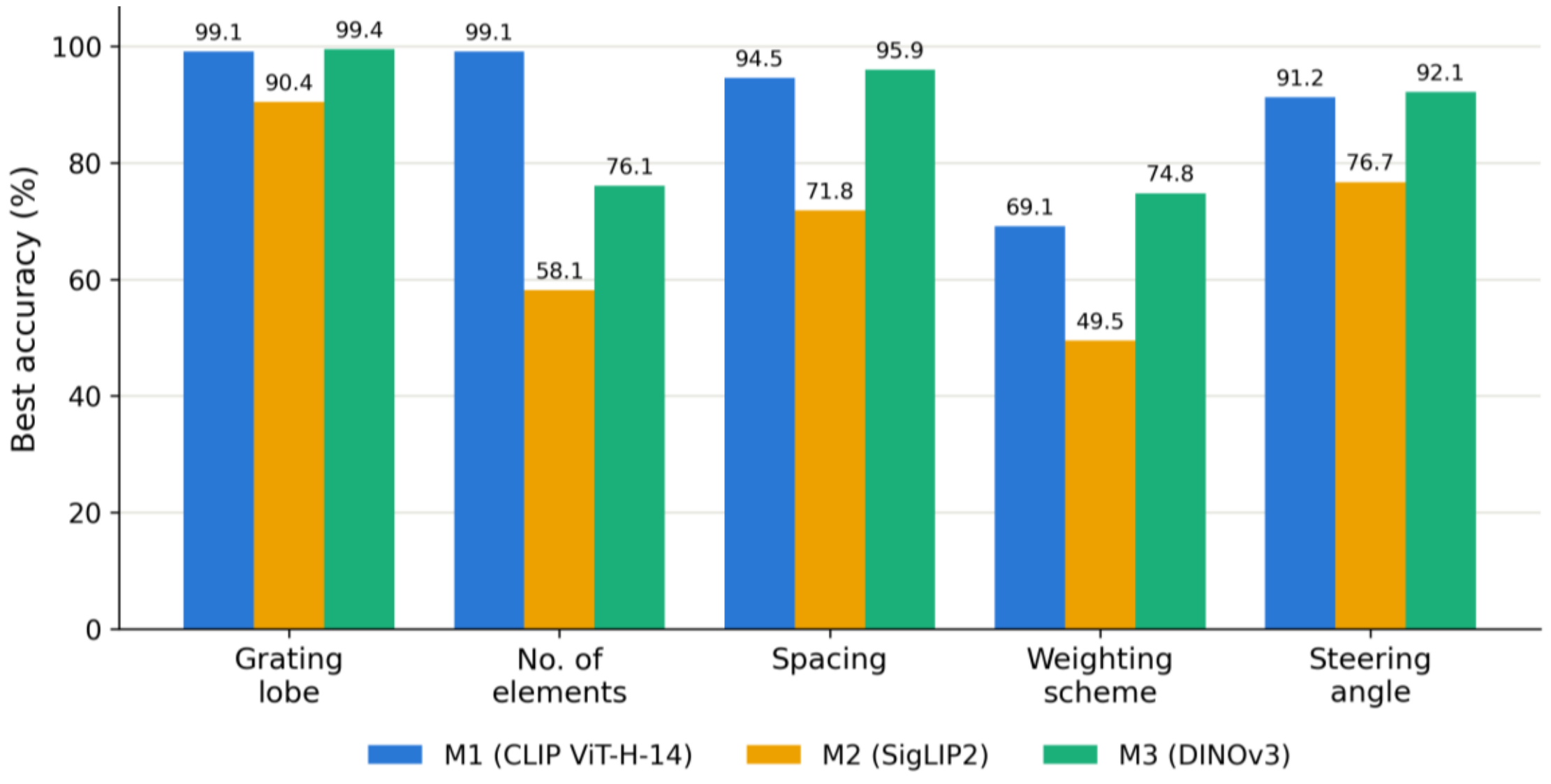

**Figure 2:** Best classification accuracy by antenna parameter and feature extractor, summarizing results in Tables II to VI.

**Weighting scheme parameter** results in Table V show that M3 with Logistic Regression gives the best accuracy of 74.79%, close to M3 with SVM (Linear) having 74.19% and M1 with SVM having ~69%. Best accuracy of M2 is only 49.5% with SVM (RBF) and goes as low as 25% with Gaussian NB. Unlike visually distinct parameters like grating lobe presence, weighting scheme is tough to classify, owing to weighting schemes (Hann, Hamming, Blackman) producing visually similar side-lobe roll-offs. that leave only few distinguishing visual cues between classes.

The results of classification for **steering angle parameter** in Table VI show that M3 with SVM (RBF) renders 92.14% accuracy, close to 91.19% of M1 with SVM (RBF) M2 favors SVM Linear with 76.7% similar to the spacing case. M2 with Gaussian NB collapses to 11.15%. Overall, M3 (dinov3) and M1 (CLIP ViT-H-14) feature extractors show consistently better performance; and across the considered classifiers SVM (RBF) yields the highest accuracy.

### *B. Results of Regression*

The regression models namely Random Forest, SVM (RBF and linear), Linear regression, kNN regression, Ridge and LASSO are assessed for their parameter prediction performance in the case of peak directivity parameter, HPBW and main lobe direction. These results are presented in Table VII to Table IX respectively using M1, M2 and M3 vision transformer feature extractors. The results are assessed using performance metrics Root Mean Square Error (RMSE), Mean Absolute Error (MAE), $R^2$ and adjusted $R^2$ values.

For **directivity parameter** (Table VII), M3 with Ridge Regression gives the lowest RMSE (4288.4) and M3 with Linear Regression gives the lowest MAE (2978.0). Ridge Regression remains best on RMSE but Linear Regression performs best on MAE and overall, M3 outperforms M1 and more significantly M2 for this parameter. For **HPBW parameter** (Table VIII), M3 with kNN achieves the best RMSE (40.771) while M3 with Linear gives lowest MAE (17.651). For **main lobe direction parameter** (Table IX), M3 with kNN performs best on both RMSE (239.43) and MAE (179.32). Notably, SVM (RBF) is unsuitable for predicting directivity and main lobe direction, as its $R^2$ values are negative or near zero.

Figure 3 provides a comparative view of the $R^2$ values across the three parameters and feature extractors. HPBW shows comparatively stronger regression performance, while directivity and main lobe direction exhibit lower $R^2$ values, indicating greater prediction difficulty. The classification results are comparatively more encouraging than the regression results, owing to regression being more sensitive than classification to the information loss introduced by image resizing and compression and further predicting precise continuous values requires finer-grained signal than separating discrete classes.

TABLE VII. REGRESSION RESULTS OF DIRECTIVITY

| **D_peak_dBi** | ViT-H-14-378-quickgeludfn5b | | | | ViT-gopt-16-SigLIP2-384webli | | | | dinov3-vit7b16-pretrain-lvd1689m | | | |
|---|---|---|---|---|---|---|---|---|---|---|---|---|
| | **RMSE** | **MAE** | **$R^2$** | **Adj. $R^2$** | **RMSE** | **MAE** | **$R^2$** | **Adj. $R^2$** | **RMSE** | **MAE** | **$R^2$** | **Adj. $R^2$** |
| Random Forest | 6051.6609 | 4713.8369 | 0.4239 | 0.3582 | 6985.1 | 5683.9 | 0.2325 | 0.0931 | 5610.7 | 4190.5 | **0.5048** | **0.1612** |
| SVM (RBF) | 9958.5634 | 6099.9152 | -0.5601 | -0.7381 | 10044.4 | 6121.1 | -0.5871 | -0.8751 | 9979.9 | 6097.6 | -0.5668 | -1.6539 |
| SVM (Linear) | 6478.1970 | 4482.8801 | 0.3398 | 0.2645 | 8629.1 | 5804.5 | -0.1714 | -0.3839 | 4310.4 | 3145.4 | 0.7077 | 0.5049 |
| Linear Regression | 5357.3212 | 3969.5397 | 0.5485 | 0.4970 | 6246.0 | 4851.0 | 0.3863 | 0.2749 | 5268.7 | **2978.0** | 0.5633 | 0.2603 |
| KNN | 5590.2263 | **3454.9905** | 0.5084 | 0.4523 | 7230.1 | 5213.3 | 0.1777 | 0.0284 | 4292.7 | 3127.0 | 0.7101 | 0.5090 |
| Ridge | **5412.6494** | 4008.4352 | 0.5391 | 0.4866 | **6237.0** | **4830.3** | 0.3881 | 0.2770 | **4288.4** | 3106.2 | 0.7107 | 0.5099 |
| LASSO | 5717.1325 | 4314.4443 | 0.4858 | 0.4272 | 6464.8 | 5006.6 | 0.3425 | 0.2232 | 5610.7 | 4190.5 | 0.5048 | 0.1612 |

TABLE VIII. REGRESSION RESULTS OF HPBW

| **HPBW** | ViT-H-14-378-quickgeludfn5b | | | | ViT-gopt-16-SigLIP2-384webli | | | | dinov3-vit7b16-pretrain-lvd1689m | | | |
|---|---|---|---|---|---|---|---|---|---|---|---|---|
| | **RMSE** | **MAE** | **$R^2$** | **Adj. $R^2$** | **RMSE** | **MAE** | **$R^2$** | **Adj. $R^2$** | **RMSE** | **MAE** | **$R^2$** | **Adj. $R^2$** |
| Random Forest | 48.5904 | 23.7766 | 0.9174 | 0.9080 | 62.141 | 35.582 | 0.8649 | 0.8404 | 43.111 | 20.600 | 0.9350 | 0.8898 |
| SVM (RBF) | 86.5093 | 37.6778 | 0.7381 | 0.7083 | 149.83 | 71.986 | 0.2144 | 0.0719 | 79.433 | 35.106 | 0.7792 | 0.6260 |
| SVM (Linear) | 54.2275 | 32.6369 | 0.8971 | 0.8854 | 96.861 | 53.737 | 0.6717 | 0.6122 | 43.564 | 29.273 | 0.9336 | 0.8875 |
| Linear Regression | 46.4110 | 31.3938 | 0.9246 | 0.9160 | 73.011 | 52.585 | 0.8135 | 0.7796 | 43.371 | **17.651** | 0.9342 | 0.8885 |
| KNN | **45.0902** | **18.6209** | 0.9289 | 0.9207 | **60.564** | **31.495** | 0.8717 | 0.8484 | **40.771** | 25.822 | 0.9418 | 0.9015 |
| Ridge | 46.9434 | 31.6296 | 0.9229 | 0.9141 | 74.060 | 52.640 | 0.8081 | 0.7733 | 52.731 | 34.480 | 0.9027 | 0.8352 |
| LASSO | 59.8587 | 40.6655 | 0.8746 | 0.8603 | 107.86 | 77.003 | 0.5929 | 0.5191 | 43.111 | 20.600 | 0.9350 | 0.8898 |

TABLE IX. REGRESSION RESULTS OF MAIN LOBE DIRECTION

| Main Lobe Direction | ViT-H-14-378-quickgeludfn5b | | | | ViT-gopt-16-SigLIP2-384webli | | | | dinov3-vit7b16-pretrain-lvd1689m | | | |
|---|---|---|---|---|---|---|---|---|---|---|---|---|
| | **RMSE** | **MAE** | **$R^2$** | **Adj. $R^2$** | **RMSE** | **MAE** | **$R^2$** | **Adj. $R^2$** | **RMSE** | **MAE** | **$R^2$** | **Adj. $R^2$** |
| Random Forest | 272.2145 | 229.7011 | 0.2847 | 0.2031 | 289.82 | 248.13 | 0.1891 | 0.0420 | 258.32 | 208.35 | 0.3558 | -0.0912 |
| SVM (RBF) | 316.0607 | 271.3187 | 0.0357 | -0.0743 | 329.57 | 287.05 | -0.0485 | -0.2389 | 312.75 | 266.31 | 0.0557 | -0.5995 |
| SVM (Linear) | 262.4200 | 198.3659 | 0.3352 | 0.2594 | 289.71 | 235.25 | 0.1898 | 0.0427 | 239.54 | 179.49 | 0.4461 | 0.0617 |
| Linear Regression | **247.6247** | **190.0544** | 0.4081 | 0.3405 | 273.34 | **218.22** | 0.2787 | 0.1478 | 266.94 | 196.12 | 0.3121 | -0.1652 |
| KNN | 275.9551 | 208.8246 | 0.2649 | 0.1810 | 306.02 | 238.69 | 0.0959 | -0.0681 | **239.43** | **179.32** | 0.4466 | 0.0625 |
| Ridge | 248.4294 | 191.6025 | 0.4042 | 0.3362 | **271.42** | 218.58 | 0.2888 | 0.1598 | 241.77 | 185.10 | 0.4357 | 0.0441 |
| LASSO | 263.2413 | 211.9086 | 0.3311 | 0.2547 | 306.12 | 272.92 | 0.0954 | -0.0688 | 258.32 | 208.35 | 0.3558 | -0.0912 |

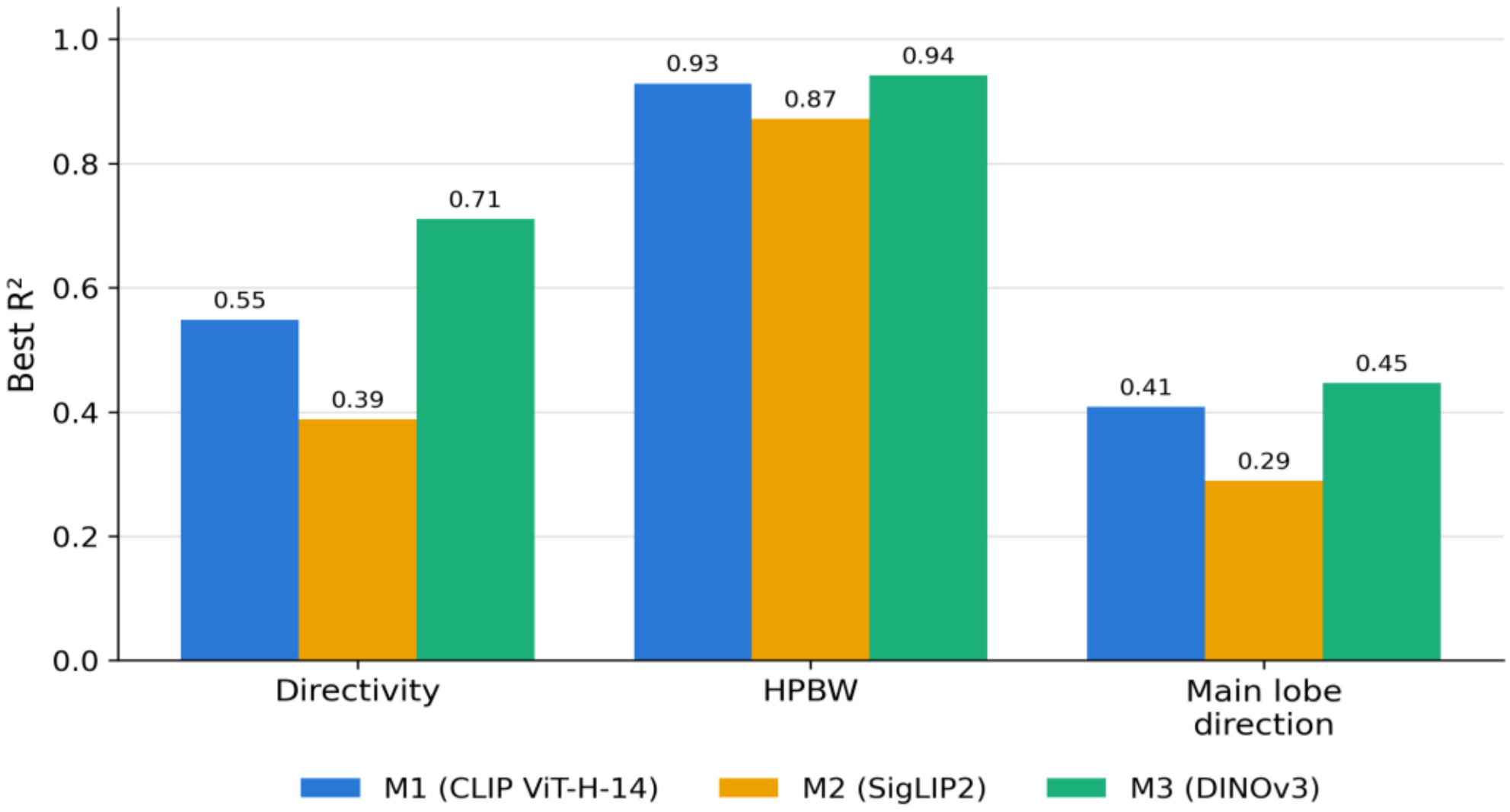

**Figure 3:** Best $R^2$ value by regression parameter and feature extractor, summarizing the $R^2$ for each parameter in Tables VII to IX

The results in Table II to Table IX demonstrate the discriminative ability of the three vision transformer models in feature extraction for radiation pattern characteristics. Overall, the combined use of high-capacity vision encoders and GPU-accelerated classical machine learning resulted in an efficient and scalable framework for analyzing synthetic antenna radiation patterns.

## VII. CONCLUSION

This work presented an AI-driven framework for automated radiation pattern interpretation, framed as a diagnostic and reverse inference problem typically relevant to OTA testing, RF surveillance, and non-cooperative emitter analysis. A large-scale synthetic dataset, RadPat-50K, comprising 50,000 radiation pattern images of ULAs with diverse design parameters and controlled perturbations, was introduced and made publicly available to promote reproducibility. High-dimensional embeddings were extracted using state-of-the-art vision transformers, capturing latent structural information directly from pattern images without explicit electromagnetic computations. Antenna parameter inference was formulated as a problem involving either classification or regression with input being a radiation pattern image, enabling estimation of discrete configuration parameters and continuous performance metrics. Rigorous train–test evaluation demonstrated consistent performance, with classification accuracies exceeding 70% and significant regression accuracy. By bridging antenna theory with modern vision–language models, this work establishes a scalable pathway for intelligent, image-based antenna diagnostics and performance inference, advancing data-driven methodologies for next-generation antenna analysis and design automation.

## COMPETING INTERESTS

The author(s) declare none.